\documentclass[aps,twocolumn,superscriptaddress]{revtex4}%
\usepackage{amsfonts}
\usepackage{amsmath}
\usepackage{amssymb}
\usepackage{graphicx}%
\providecommand{\U}[1]{\protect\rule{.1in}{.1in}}

\begin{document}
\preprint{ LLNL-JRNL-401466}
\title[Temperature Equilibration]{Molecular Dynamics Simulations of 
Temperature Equilibration in Dense Hydrogen}
\author{J. N. Glosli}
\affiliation{Lawrence Livermore National Laboratory}
\author{F. R. Graziani}
\affiliation{Lawrence Livermore National Laboratory}
\author{R. M. More}
\affiliation{Lawrence Livermore National Laboratory}
\author{M. S. Murillo}
\affiliation{Physics Division, MS D410, Los Alamos National 
Laboratory, Los Alamos, NM 87545}
\author{F. H. Streitz}
\affiliation{Lawrence Livermore National Laboratory}
\author{M.~P.~Surh}
\affiliation{Lawrence Livermore National Laboratory}
\author{L. X. Benedict}
\affiliation{Lawrence Livermore National Laboratory}
\author{S. Hau-Riege}
\affiliation{Lawrence Livermore National Laboratory}
\author{A. B. Langdon}
\affiliation{Lawrence Livermore National Laboratory}
\author{R. A. London}
\affiliation{Lawrence Livermore National Laboratory}

\begin{abstract}
The temperature equilibration rate in dense hydrogen (for both $T_{i}>T_{e}$
and $T_{i}<T_{e}$) has been calculated with molecular dynamics
simulations for temperatures between 10 and 600 eV and densities between
$10^{20}/cc$ to $10^{24}/cc$. Careful attention has been devoted to
convergence of the simulations, including the role of semiclassical
potentials. We find that for Coulomb logarithms 
$\mathcal{L}\gtrsim1$, a model by Gericke-Murillo-Schlanges (GMS) 
[Gericke et al., PRE {\bf 65}, 036418 (2002)]
based on a T-matrix method and the approach by 
Brown-Preston-Singleton [Brown et al., Phys. Rep. \textbf{410}, 237
(2005)] agrees with the simulation data to within the error bars of 
the simulation. For smaller Coulomb
logarithms, the GMS model is consistent with the simulation results. 
Landau-Spitzer models are
consistent with the simulation data for ${\cal L}>4$.

\end{abstract}
\date{\today}
\received{}

\startpage{1}
\endpage{2}
\maketitle
\endpage{ }

\DeclareGraphicsExtensions{.pdf,.jpg,.png,.gif}

\section{Introduction}

The strong temperature dependence of thermonuclear reaction rates suggests that
even small deviations from equilibrium can yield differences in burn rates.
Thus, the pursuit of ignition in the laboratory will benefit from 
accurate models of
relaxation processes in hot, dense plasmas.
One of the greatest uncertainties in the nonequilibrium energy balance is the
electron-ion temperature relaxation rate.  Although there have been indirect
measurements for cool dense matter \cite{1}, there is no experimental 
data in the regime
of interest.  Even worse, theoretical descriptions of Coulomb 
collisions suffer from
divergences that make detailed models difficult to develop. Here we take a complementary approach to hot,
dense plasmas by using molecular dynamics (MD) techniques. We use 
this method to test recent theoretical models and compare with 
standard results.

The electron-proton coupling rate was first calculated by Landau \cite{2}
and Spitzer (LS) \cite{3} for classical plasmas with weak collisions.
They write the electron-proton temperature exchange rate $\left( 
1/\tau_{pe}\right)  $ in the form,%
\begin{equation}
\frac{1}{\tau_{pe}}=\frac{8\sqrt{2\pi}n_{p}Z^2e^{4}}{3m_{e}m_{p}c^{3}%
}\left\{ 
\frac{k_BT_{e}}{m_{e}c^{2}}+\frac{k_BT_{p}}{m_{p}c^{2}}\right\} 
^{-3/2}%
\mathcal{L}\equiv\frac{\mathcal{L}}{J_{LS}},
\end{equation}
where $J_{LS}$ is the LS pre-factor, $n_{e}$ ($n_{p}$) are the electron
(ion) number densities, $Z=1$ is the proton charge, $T_{e}$ ($T_{p}$)
are the electron (ion) temperatures, and $k_B$ is the Boltzmann constant.
$\mathcal{L}$ is the so-called Coulomb logarithm containing details 
of the collision process. 
LS used
\begin{equation}
{\cal L}_{LS}=\ln(b_{max}/b_{min})
\end{equation}
where $b_{max}$ and $b_{min}$ are impact parameter
cutoffs needed to remove divergences that arose from their treatment. 
$b_{min}$ is chosen to be a
minimum impact parameter consistent with plasma conditions, such as 
the classical distance of closest approach ($b_C = Ze^2/k_BT$). At 
high temperatures, $b_{min}$ is
often modified to include quantum diffraction effects by introducing 
the length scale of the
electron thermal deBroglie wavelength 
$\Lambda=\sqrt{2\pi\hslash^{2}/m_{e}k_BT_{e}}$.  Typically $b_{\max}$ 
is chosen to
be a screening length arising from collective plasma phenomena, such 
as  the  Debye length $\lambda_{D}=\sqrt{k_BT_{e}/4\pi e^{2}n_{e}}$.

The presence of {\it ad hoc} cut-offs and other inconsistencies led researchers to
derive kinetic equations without cut-offs \cite{5,6,6a,7}.
The essence of these theories is the inclusion of strong scattering 
in the presence of dynamical
collective (screening) behavior. Two such theories have been recently proposed:  Gericke, Murillo and 
Schlanges (GMS) \cite{7} and Brown, Preston and Singleton (BPS) \cite{8}.
GMS  applied these
ideas to dense plasma temperature equilibration, They investigated various
approximations in evaluating of $\mathcal{L}$, including issues with 
trajectories and cutoffs, and provided four different evaluations of 
the relaxation
rate based on quantum kinetic theory. From their numerical work, GMS 
suggest an effective Coulomb logarithm \cite{8b}
\begin{equation}
\mathcal{L}_{GMS6}=\frac{1}{2}\ln\left(  1+\left[ 
\lambda_{D}^{2}+R_{ion}^{2}\right]  /\left[
\Lambda^{2}/8\pi+b_{C}^{2}\right]  \right),
\end{equation}
where $R_{ion}=\left(  3/4\pi n_{p}\right)  ^{1/3}$ is the ion sphere 
radius. This expression was described by GMS as the best fit to their full T-matrix theory.

BPS and Brown and Singleton
\cite{9} used dimensional continuation to obtain an expression for the
electron-ion coupling rate accurate to second order in the plasma coupling parameter. The method
is applicable to both degenerate and non-degenerate electrons. For the non-degenerate case, they derive 

\begin{equation}
\mathcal{L}_{BPS}=\log(\lambda_{D}/\Lambda)+\left(  \log\left(  16\pi\right)
-\gamma-1\right)  /2,
\end{equation}
where $\gamma$ is the Euler constant.

The most direct method of studying temperature equilibration in the classical 
limit is with numerical
simulation; strong, collective scattering at all length scales is the 
forte of MD. Hansen and McDonald (HM)
\cite{10} explored temperature equilibration in dense hydrogen  using MD,
comparing their results against a LS model with ${\cal L}_{LS}=\ln(2\pi\lambda_{D}/\Lambda)$. 
However, the HM simulations involved a very small number
of particles $\left(  N=128\right) $ with presumably large error bars.
Here, we expand upon their calculations to
not only reassess the HM result, but also compare with the modern 
approaches of GMS and BPS.

\section{Molecular Dynamics: Simulations and Results}

MD simulations are applied to two-temperature systems of charged particles in a
cubic cell with periodic boundary conditions. The MD is performed with a fully parallel code using a
basic leapfrog method\cite{11} with the
Coulomb interaction evaluated by an Ewald summation 
\cite{12,12a}.
Because the classical Coulomb many-body problem is unstable for attractive
interactions, we employ semi-classical potentials that reduce
the Coulomb interaction on short length scales in order to
prevent unphysical, deeply-bound states.
We tested several forms of the diffractive \cite{14,15} and Pauli 
\cite{16,17} terms for  these potentials.
The resulting equilibration times typically vary by less than 15\%, which is within 
the statistical error of the MD data.
The similarity is not unexpected, since
most semi-classical potentials resemble one another above $10$ eV 
\cite{filinov}.
We report
results using the semi-classical potential in HM \cite{10},
\begin{eqnarray}
\label{DB}
V_{ab}(r)  &=&\frac{Z_a Z_b e^2}{r}\left( 1-\exp\left(
-2\pi r/\Lambda_{ab}\right)  \right) \,\cr
 &+&k_BT\ln2  \exp\left(
-4\pi  r^2/\Lambda^2_{ab}/\ln2\right) \delta_{ae}\,\delta_{be}
\end{eqnarray}
where $\Lambda_{ab}=\sqrt{2\pi
\hbar^{2}/\mu_{ab}k_BT}$, $\mu_{ab}$ is the reduced mass, and $T=T_{e}$
except when $a$ and $b$ are both protons when $T=T_{p}$. The potentials are
temperature-dependent, but were held constant in most of our short 
simulations. 
For long simulations to equilibration, we allow the 
temperature parameters to evolve with time, using a smoothed 
exponential average of the instantaneous MD value.

Simulations were run long enough to extract a relaxation time 
(typically 10\% of
$\tau^*$), with some strong-coupling cases continued to complete
equilibration. We obtain $\tau_{pe}$ by fitting the temperature over a brief interval,
\begin{equation}
\frac{dT_{e}}{dt}=\frac{T_{p}-T_{e}}{\tau_{ep}};\qquad
\frac{dT_{p}}{dt}=\frac{T_{e}-T_{p}}{\tau_{pe}}.
\label{five}
\end{equation}
We choose
the timestep to conserve total energy over the entire simulation
$\left(  \Delta E/E<10^{-4}\right)$ when using fixed 
potentials.
Typically, $\Delta t$  ranges from $5\times10^{-5}$ to
$10^{-3}$ fs.
Any drift in the energy is tightly controlled, as artificial heating 
can distort the true relaxation rate.
In long runs, the potentials change slowly as the temperature 
relaxes. 
Although energy is not conserved in these cases, the total energy change remains less than 3\%.  In practice, 
$\tau_{pe}$ calculated from the time-dependent potential is within 
10-15\% of the result for the constant potential.

Convergence with 
respect to system size is tested by employing various
particles numbers N ranging from $N=128$ (the number that HM employed), to as
many as $N=64,000$.
The results reported here use $N=1024$.
Statistical uncertainty for 
each case is estimated
by computing the relaxation rate from 
equivalent samples (from 8 to 64) of a microcanonical ensemble  and 
then taking the average and standard deviation. 
Sensitivity to 
initial conditions is studied using the ensemble of simulations 
and/or by discarding a
portion of the initial temperature evolution.

The nonequilibrium system is prepared using two separate Langevin
thermostats for protons and electrons. Initial configurations are sampled from
a stationary distribution obtained after $10^{5}-10^{7}$ timesteps. The
thermostats are then removed, and the species allowed to undergo 
(microcanonical) collisional
relaxation for approximately $10^6$ timesteps. 
  
Equations \ref{five} are valid for an ideal gas equation of state for the plasma.
For strongly coupled plasmas there is a significant potential energy 
contribution that would invalidate this assumption. However, the error associated with using the temperature evolution equations is small in the temperature-density regimes of interest here\cite{13}.
Although the MD temperature relaxation is asymmetric in the
strong-coupling cases, we find $\left\vert
dT_{e}/dt\right\vert $ and $\left\vert dT_{p}/dt\right\vert $ differ by only
about 10\%. Thus, we only report $1/\tau^*=1/\tau_{pe}+1/\tau_{ep} \approx 2/\tau_{pe}$.

\begin{table}[ptb]
\hfil%
\begin{tabular}
[c]{cccccc}\hline\hline
Case & $n_{i}(1/cc)$ & $T_{e}(eV)$ & $T_{i}(eV)$ & $\tau^* (fs)$
& $\sigma(fs)$\\\hline
A & $10^{20}$ & $10.0$ & $20.0$ & $2.04\times10^{4}$ & $4.9\times10^{3}$\\
B & $10^{20}$ & $30.0$ & $60.0$ & $7.89\times 10^4$& $4.3 \times 10^4$\\
C & $10^{22}$ & $10.0$ & $20.0$ & $5.23 \times 10^2$ & $1.7\times 10^2$\\
D & $10^{22}$ & $30.0$ & $60.0$ & $1.73\times10^{3}$ & $6.6\times10^{2}$\\
E & $10^{22}$ & $100.0$ & $200.0$ & $6.45\times10^{3}$ & $2.2\times10^{3}$\\
F & $10^{24}$ & $10.0$ & $20.0$ & $8.87\times10^{1}$ & $3.5\times10^{1}$\\
G & $10^{24}$ & $30.0$ & $60.0$ & $8.27\times10^{1}$ & $3.3\times10^{1}$\\
H & $10^{24}$ & $100.0$ & $200.0$ & $1.72\times10^{2}$ & $6.2\times10^{1}$\\
I & $10^{24}$ & $300.0$ & $600.0$ & $4.17\times10^{2}$ & $8.0\times10^{1}$\\
J & $1.61\times10^{24}$ & $29.9$ & $80.1$ & $20.2\times10^{1}$ & $5.3$\\
K & $1.61\times10^{24}$ & $91.47$ & $12.1$ & $1.20\times10^{2}$ &$1.7\times10^1$\\ \hline
L& $10^{20}$ & $100.0$ & $200.0$ & $3.65 \times 10^5$ & $3.2\times 10^5$\\
M$_1$ & $10^{20}$ & $10.0$ & $40.0$ & $2.05\times10^{4}$ 
&$3.0\times10^3$\\  
M$_2$ & $10^{20}$ & $10.0$ & $40.0$ & $2.18\times10^{4}$ 
&$4.5\times10^3$\\                   
M$_3$ & $10^{20}$ & $10.0$ & $40.0$ & $2.28\times10^{4}$ 
&$9.6\times10^3$\\  
\hline\hline
\end{tabular}
\hfil\caption{Density, initial electron, and ion temperature, relaxation time
and standard deviation of the MD simulations. }%
\end{table}

Table 1 lists the set of initial conditions for 15 different 
systems.
The ensemble average temperature relaxation, $\tau^*$ 
(calculated from $d\Delta T/dt = \Delta T/\tau^*$),
and the 
standard deviation, $\sigma$, are in femtoseconds.
A range of initial 
conditions were chosen to span the weakly- to strongly-coupled and the degenerate to
non-degenerate regimes. 
We include two sets of initial conditions 
considered by HM (Cases J and K).
In most cases, hydrogen plasma is 
simulated using the true electron-proton mass ratio of
1:1836. 
In Case L, the cold electrons were replaced with cold 
protons in order to shorten the required simulation time. 
Cases 
M$_{1-3}$ involve a comparison of 
electron-proton and 
positron-proton systems and will be discussed below.  
Cases F and G 
have degenerate electrons.  Degeneracy effects are treated in neither
the classical MD simulations nor in the LS,  GMS6,
or  BPS models.   Hence, the models can be directly compared to the 
simulations even for those cases when
comparisons with experiment would be questionable.

\section{Comparison with Theory}

\begin{figure}
[ptb]
\begin{center}
\includegraphics[width=\columnwidth]{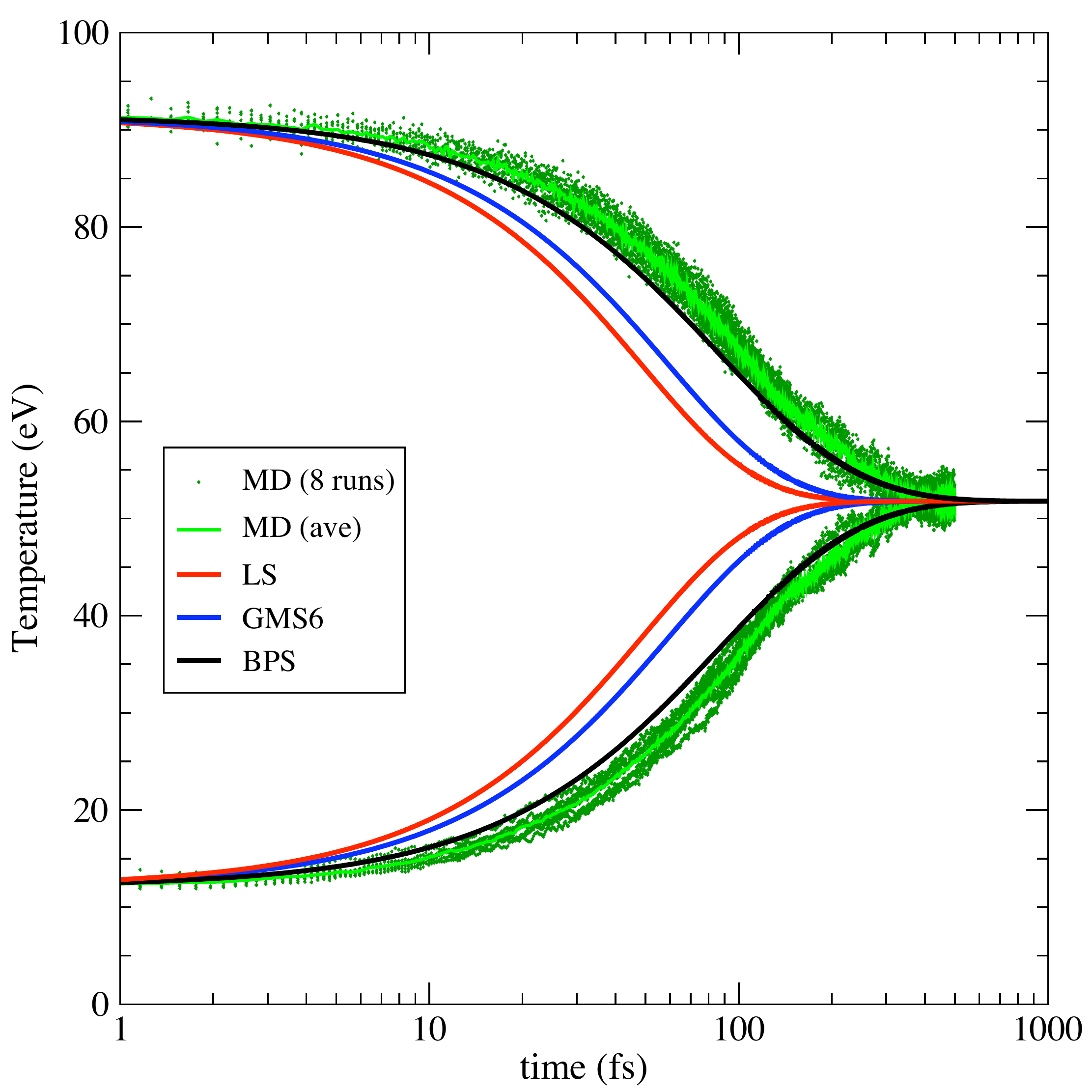}%
\caption{\label{fig1} Electron (top curves) and proton (bottom 
curves) temperature relaxation is shown based on
MD, GMS, LS, and BPS for Case L. The MD results are shown by points 
from several simulations, with a line
through the average. Note that all approaches relax slower than LS.}%
\label{Figure 1}%
\end{center}
\end{figure}

Figure \ref{fig1} shows MD results for Case K run to near-full 
relaxation using potentials that are implicitly time-dependent 
(temperature-dependent).  We also display predictions for LS, GMS6, 
and BPS.  The MD data in
Figure \ref{fig1} is most closely matched by BPS (although this is partly fortuitous, as will been 
seen below) followed by GMS6.
The LS model predicts the fastest relaxation, exceeding MD 
by about a factor of two. 
This disagreement  contradicts the 
conclusion reached by HM.
At the same time, our $\tau^*$ for Cases J and K agree with 
those reported by HM. We attribute the discrepancy to 
inconsistent definitions of $\tau_{pe}$, $\tau_{LS}$ and $\tau^*$: $\tau_{LS}$ is properly equal to $\tau_{pe}$, which is $2\tau^*$ (not $\tau^*$) if  $\tau_{pe}\equiv\tau_{ep}$. As previously noted by HM, however, ambiguities in the $b_{min}$ and $b_{max}$ may be 
sufficient to accommodate this difference.

\begin{figure}
[ptb]
\begin{center}
\includegraphics[width=\columnwidth]{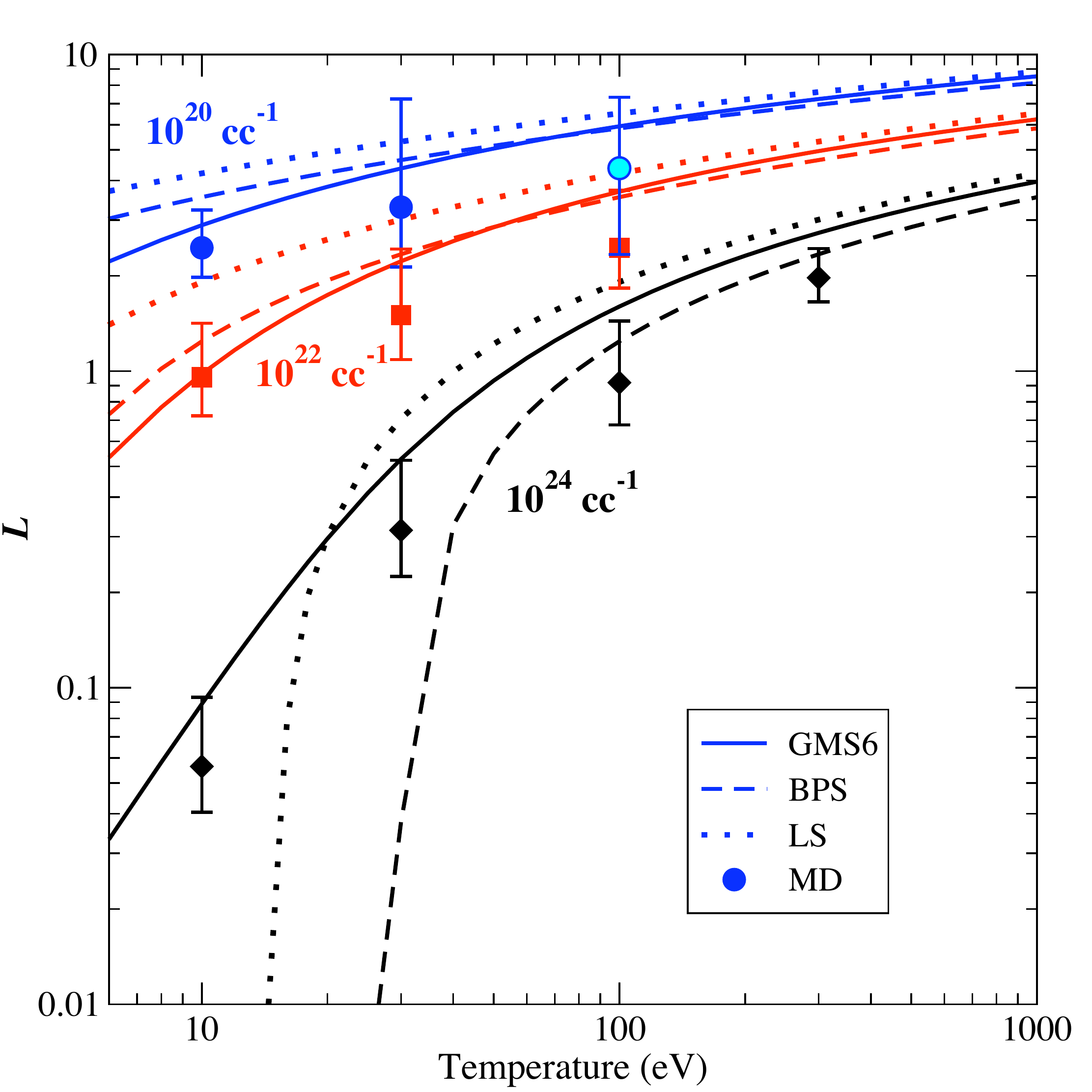}%
\caption{\label{fig2} Theoretical (GMS6 [solid], BPS [dashed], and LS 
[dotted]) and MD
calculations of $\mathcal{L}$ as a function of initial $T_{e}$ for densities
$10^{20}/cc,10^{22}/cc,$ and $10^{24}/cc$ (blue, red and black
respectively). Additional detail is in the text.}%
\label{Figure 2}%
\end{center}
\end{figure}

To make comparisons of our MD results with theoretical predictions 
more transparent, we define an
effective Coulomb logarithm as ${\cal L}_{MD}\equiv 2 
J_{LS}/\tau^*$.  This result is then
compared with the theoretical prediction for $\mathcal{L}$ coming 
from LS, GMS6, and BPS. Fig. (\ref{fig2}) shows simulation results 
for $\mathcal{L}_{MD}$ with error bars along with theoretical 
predictions for $\mathcal{L}_{GMS6}$ (solid) and $\mathcal{L}_{BPS}$ 
(dashed) as a function of initial electron temperature. 
Numerical 
results and analytic expressions for $\mathcal{L}$ are arranged 
according to density; $n=10^{20},10^{22},$ and $10^{24}$ (blue,
red and black respectively).

In regions where it is expected to be applicable, we find that LS 
systematically overestimates
the effective Coulomb logarithm and thus predicts a relaxation rate 
that is too fast
relative to the MD results.
For plasmas with $\mathcal{L}>1$, the MD results are consistent with 
both the GMS6 and BPS, suggesting that approaches beyond LS are 
indeed more predictive.  As expected, BPS increasingly underestimates 
the relaxation rate for  ${\cal L}<1$; BPS is not intended for use in 
this regime.    
For the case shown in Fig.~\ref{fig1}, the 
underestimation  at lower temperatures compensates
for an overestimation at early times, making  agreement with this 
simulation fortuitously good. As is evident from Figure \ref{fig2}, 
this would not be the case in general \cite{foot}. We find that GMS6 
captures the qualitative variation of $\cal L$ over a surprisingly 
broad range of density and temperature.  Further discrimination 
between these theories in the region where where they are expected to 
be most accurate (low density and high temperature) is not possible 
given the large uncertainties present in our current MD simulations. 
However, our results suggest that validation of these theories could 
be accomplished with carefully controlled experiments \cite{19} and 
larger (and longer) simulations that further reduce statistical error.

Finally, LS predicts identical equilibration rates for like-charge 
and opposite charge systems. We tested this by performing three sets 
of simulations at the same density and temperatures (Case M$_{1-3}$ 
in Table I.)  We simulated electrons-protons (M$_1$) and 
positrons-protons (M$_2$) using Equation 6, and positrons-protons 
using a pure $1/r$ Coulomb potential (M$_3$). The relaxation rates 
for all three cases agree to within our error bars, suggesting that 
energy transfer in these systems is occurring predominately on length 
scales longer than the thermal deBroglie wavelength.

\section{Conclusions}

We have performed MD simulations  of the temperature relaxation process in hot, dense 
hydrogen.  We investigated systems containing as large as 64,000 
particles,  finding that $N\simeq$1000 particles is sufficient for most cases we 
considered. Our simulations span a large range of temperature and 
density parameter space, including the first simulations in the 
low-density, high-temperature limit.  

For the
weakly coupled plasmas where $\mathcal{L}\gtrsim1$, the simulations 
are consistent with both GMS6 and BPS.  In contrast,  the LS approach 
systematically
overestimates the relaxation rate.  In the limit of high temperature and
low density, all models are in agreement, however.  Our MD results 
suggest that LS is accurate for ${\cal L} > 4$, rather than the
usual restriction of ${\cal L}>10$, in agreement with previous work 
\cite{7,18}.
More modern approaches exemplified here by GMS6 and BPS clearly extend the accessible 
parameter space closer
to ${\cal L}\sim 1$, with GMS6 providing a reasonable description of the MD data even for
for ${\cal L}<1$.

We have employed two forms of the semiclassical potentials needed for stability in an MD simulation with attractive potentials, and have found a very slight effect from the form of the potential; as such, we believe that our results are not sensitive to the choice of the semiclassical portion of the potentials.

\section{Acknowledgments}

This work performed under the auspices of the U.S. Department of Energy by
Lawrence Livermore National Laboratory under Contract DE-AC52-07NA27344. This
work was funded by the Laboratory Directed Research and Development Program at
LLNL under project tracking code 07-ERD-044.


\begin{thebibliography}{99}
\bibitem {8b} This is model 6 in Ref. \cite{7}. Note that we have 
recast Equations 3 and 4 to use a consistent definition of the 
thermal deBroglie wavelength.
\end{thebibliography}
\end{document}